\documentclass{aastex62}

\usepackage{xspace}
\newcommand{\starname}{LP 791-18} 
\newcommand{\TESS}{{\em TESS}} 

\received{\today}
\submitjournal{ApJL}

\shorttitle{Two planets transiting \starname}
\shortauthors{Crossfield et al.}

\begin{document}

\title{A super-Earth and sub-Neptune transiting the late-type M dwarf \starname}

\correspondingauthor{Ian J.\ M.\ Crossfield}
\email{iancross@mit.edu}

\author{Ian J.\ M.\ Crossfield}
\affiliation{Department of Physics, and Kavli Institute for Astrophysics and Space Research, Massachusetts Institute of Technology, Cambridge, MA, USA}
\nocollaboration

\author{William Waalkes}
\altaffiliation{NSF Graduate Research Fellow}
\affiliation{Department of Astrophysical and Planetary Sciences, University of Colorado, Boulder, CO, USA}
\author[0000-0003-4150-841X]{Elisabeth R. Newton}
\affiliation{Department of Physics and Astronomy, Dartmouth College, Hanover, NH, USA}
\author{Norio Narita}
\affiliation{ Astrobiology Center, 2-21-1 Osawa, Mitaka, Tokyo 181-8588, Japan}
\affiliation{ JST, PRESTO, 2-21-1 Osawa, Mitaka, Tokyo 181-8588, Japan}
\affiliation{ National Astronomical Observatory of Japan, 2-21-1 Osawa, Mitaka, Tokyo 181-8588, Japan}
\affiliation{ Instituto de Astrof\'{i}sica de Canarias (IAC), 38205 La Laguna, Tenerife, Spain}
\author{Philip Muirhead}
\affiliation{Department of Astronomy \& Institute for Astrophysical Research, Boston University, 725 Commonwealth Ave., Boston, MA, USA}
\author{Kristo Ment}
\affiliation{Center for Astrophysics,  Harvard {\rm \&} Smithsonian, 60 Garden Street, Cambridge, MA, USA}
\author{Elisabeth Matthews}
\affiliation{Department of Physics, and Kavli Institute for Astrophysics and Space Research, Massachusetts Institute of Technology, Cambridge, MA, USA}
\author{Adam Kraus}
\affiliation{Department of Astronomy, University of Texas at Austin, Austin, TX, USA}
\author{Veselin Kostov}
\affiliation{NASA Goddard Space Flight Center, 8800 Greenbelt Road, MD, USA}
\author{Molly R.\ Kosiarek}
\altaffiliation{NSF Graduate Research Fellow}
\affiliation{Department of Astronomy and Astrophysics, University of California, Santa Cruz, CA, USA}
\author{Stephen R.\ Kane}
\affiliation{Department of Earth and Planetary Sciences, University of California, Riverside, CA, USA}
\author{Howard Isaacson}
\affiliation{Astronomy Department, University of California, Berkeley, CA, USA}
\author{Sam Halverson}
\affiliation{Department of Physics, and Kavli Institute for Astrophysics and Space Research, Massachusetts Institute of Technology, Cambridge, MA, USA}
\author{Erica Gonzales}
\altaffiliation{NSF Graduate Research Fellow}
\affiliation{Department of Astronomy and Astrophysics, University of California, Santa Cruz, CA 95064, USA}
\author{Mark Everett}
\affiliation{National Optical Astronomy Observatory, Tucson, AZ, USA}
\author{Diana Dragomir}
\affiliation{Department of Physics, and Kavli Institute for Astrophysics and Space Research, Massachusetts Institute of Technology, Cambridge, MA, USA}
\author[0000-0001-6588-9574]{Karen A.\ Collins}
\affiliation{Center for Astrophysics,  Harvard {\rm \&} Smithsonian, 60 Garden Street, Cambridge, MA, USA}
\author[0000-0003-1125-2564]{Ashley Chontos}
\altaffiliation{NSF Graduate Research Fellow}
\affiliation{Institute for Astronomy, University of Hawai`i, 2680 Woodlawn Drive, Honolulu, HI, USA}
\author{David Berardo}
\altaffiliation{NSERC Graduate Research Fellow}
\affiliation{Department of Physics, and Kavli Institute for Astrophysics and Space Research, Massachusetts Institute of Technology, Cambridge, MA, USA}
\nocollaboration

\author[0000-0001-6031-9513]{Jennifer G.\ Winters}
\affiliation{Center for Astrophysics,  Harvard {\rm \&} Smithsonian, 60 Garden Street, Cambridge, MA, USA}
\author{Joshua N.\ Winn}
\affiliation{Princeton University, Department of Astrophysical Sciences, 4 Ivy Lane, Princeton, NJ, USA}
\author[0000-0003-1038-9702]{Nicholas J.\ Scott}
\affiliation{NASA Ames Research Center, Moffett Field, CA, USA}
\author{Barbara Rojas-Ayala}
\affiliation{Departamento de Ciencias Fisicas, Universidad Andres Bello, Fernandez Concha 700, Las Condes, Santiago, Chile}
\author{Aaron C.\ Rizzuto}
\altaffiliation{51 Pegasi b Fellow}
\affiliation{Department of Astronomy, University of Texas at Austin, Austin, TX, USA}
\author{Erik A.\ Petigura}
\affiliation{Department of Astronomy, California Institute of Technology, Pasadena, CA, USA}
\author{Merrin Peterson}
\affiliation{Departement de Physique, and Institute for Research on Exoplanets, Universite de Montreal, Montreal, H3T J4, Canada}
\author{Teo Mocnik}
\affiliation{Department of Earth and Planetary Sciences, University of California, Riverside, CA, USA}
\author{Thomas Mikal-Evans}
\affiliation{Department of Physics, and Kavli Institute for Astrophysics and Space Research, Massachusetts Institute of Technology, Cambridge, MA, USA}
\author{Nicholas Mehrle}
\affiliation{Department of Physics, and Kavli Institute for Astrophysics and Space Research, Massachusetts Institute of Technology, Cambridge, MA, USA}
\author[0000-0001-7233-7508]{Rachel Matson}
\affiliation{NASA Ames Research Center, Moffett Field, CA, USA}
\author{Masayuki Kuzuhara}
\affiliation{ Astrobiology Center, 2-21-1 Osawa, Mitaka, Tokyo 181-8588, Japan}
\affiliation{ National Astronomical Observatory of Japan, 2-21-1 Osawa, Mitaka, Tokyo 181-8588, Japan}
\author{Jonathan Irwin}
\affiliation{Center for Astrophysics,  Harvard {\rm \&} Smithsonian, 60 Garden Street, Cambridge, MA, USA}
\author[0000-0001-8832-4488]{Daniel Huber}
\affiliation{Institute for Astronomy, University of Hawai`i, 2680 Woodlawn Drive, Honolulu, HI, USA}
\author{Chelsea Huang}
\affiliation{Department of Physics, and Kavli Institute for Astrophysics and Space Research, Massachusetts Institute of Technology, Cambridge, MA, USA}
\author{Steve Howell}
\affiliation{NASA Ames Research Center, Moffett Field, CA, USA}
\author{Andrew W.\ Howard}
\affiliation{Department of Astronomy, California Institute of Technology, Pasadena, CA, USA}
\author{Teruyuki Hirano}
\affiliation{ Department of Earth and Planetary Sciences, Tokyo Institute of Technology, 2-12-1 Ookayama, Meguro-ku, Tokyo 152-8551, Japan}
\author{Benjamin J.\ Fulton}
\affiliation{NASA Exoplanet Science Institute, California Institute of Technology, M/S 100-22, 770 S. Wilson Ave, Pasadena, CA, USA}
\author[0000-0001-9823-1445]{Trent Dupuy}
\affiliation{Gemini Observatory, Northern Operations Center, 670 N. Aohoku Place, Hilo, HI, USA}
\author{Courtney D.\ Dressing}
\affiliation{Astronomy Department, University of California, Berkeley, CA, USA}
\author{Paul A.\ Dalba}
\affiliation{Department of Earth and Planetary Sciences, University of California, Riverside, CA, USA}
\author{David Charbonneau}
\affiliation{Center for Astrophysics,  Harvard {\rm \&} Smithsonian, 60 Garden Street, Cambridge, MA, USA}
\author{Jennifer Burt}
\affiliation{Department of Physics, and Kavli Institute for Astrophysics and Space Research, Massachusetts Institute of Technology, Cambridge, MA, USA}
\author[0000-0002-3321-4924]{Zachory Berta-Thompson}
\affiliation{Department of Astrophysical and Planetary Sciences, University of Colorado, Boulder, CO, USA}
\author{Bj\"orn Benneke}
\affiliation{Departement de Physique, and Institute for Research on Exoplanets, Universite de Montreal, Montreal, H3T J4, Canada}
\nocollaboration

\author{Noriharu Watanabe}
\affiliation{ National Astronomical Observatory of Japan, 2-21-1 Osawa, Mitaka, Tokyo 181-8588, Japan}
\affiliation{Department of Astronomical Science, Graduate University for Advanced Studies (SOKENDAI), Mitaka, Tokyo 181-8588, Japan}
\author[0000-0002-6778-7552]{Joseph D.\ Twicken}
\affiliation{NASA Ames Research Center, Moffett Field, CA, USA}
\affiliation{SETI Institute, 189 N. Bernardo No.\ 200, Mountain View, CA, USA }
\author{Motohide Tamura}
\affiliation{ Astrobiology Center, 2-21-1 Osawa, Mitaka, Tokyo 181-8588, Japan}
\affiliation{ National Astronomical Observatory of Japan, 2-21-1 Osawa, Mitaka, Tokyo 181-8588, Japan}
\affiliation{Department of Astronomy, University of Tokyo, 7-3-1 Hongo, Bunkyo-ku, Tokyo 113-0033, Japan}
\author{Joshua Schlieder}
\affiliation{NASA Goddard Space Flight Center, 8800 Greenbelt Road, MD, USA}
\author[0000-0002-6892-6948]{S.~Seager}
\affiliation{Department of Physics, and Kavli Institute for Astrophysics and Space Research, Massachusetts Institute of Technology, Cambridge, MA, USA}
\affiliation{Department of Earth, Atmospheric and Planetary Sciences, Massachusetts Institute of Technology, Cambridge, MA, USA}
\affiliation{Department of Aeronautics and Astronautics, MIT, 77 Massachusetts Avenue, Cambridge, MA, USA}
\author{Mark E.\ Rose}
\affiliation{NASA Ames Research Center, Moffett Field, CA, USA}
\author{George Ricker}
\affiliation{Department of Physics, and Kavli Institute for Astrophysics and Space Research, Massachusetts Institute of Technology, Cambridge, MA, USA}
\author{Elisa Quintana}
\affiliation{NASA Goddard Space Flight Center, 8800 Greenbelt Road, MD, USA}
\author{S\'ebastien L\'epine}
\affiliation{Department of Physics and Astronomy, Georgia State University, GA, USA}
\author[0000-0001-9911-7388]{David W.\ Latham}
\affiliation{Center for Astrophysics,  Harvard {\rm \&} Smithsonian, 60 Garden Street, Cambridge, MA, USA}
\author{Takayuki Kotani}
\affiliation{ Astrobiology Center, 2-21-1 Osawa, Mitaka, Tokyo 181-8588, Japan}
\affiliation{ National Astronomical Observatory of Japan, 2-21-1 Osawa, Mitaka, Tokyo 181-8588, Japan}
\author{Jon M.\ Jenkins}
\affiliation{NASA Ames Research Center, Moffett Field, CA, USA}
\author{Yasunori Hori}
\affiliation{ Astrobiology Center, 2-21-1 Osawa, Mitaka, Tokyo 181-8588, Japan}
\affiliation{ National Astronomical Observatory of Japan, 2-21-1 Osawa, Mitaka, Tokyo 181-8588, Japan}
\author{Knicole Colon}
\affiliation{NASA Goddard Space Flight Center, 8800 Greenbelt Road, MD, USA}
\author{Douglas A.\ Caldwell}
\affiliation{SETI Institute, 189 N. Bernardo No.\ 200, Mountain View, CA, USA }


\begin{abstract}
Planets occur most frequently around cool dwarfs, but only a handful
of specific examples are known to orbit the latest-type M stars. Using
{\em TESS} photometry, we report the discovery of two planets
transiting the low-mass star called LP 791-18 (identified by {\em
  TESS} as TOI 736). This star has spectral type M6V, effective
temperature 2960 K, and radius 0.17$R_\odot$, making it the
third-coolest star known to host planets. The two planets straddle the
radius gap seen for smaller exoplanets; they include a $1.1 R_\oplus$
planet on a 0.95~day orbit and a $2.3 R_\oplus$ planet on a 5~day
orbit. Because the host star is small the decrease in light during these
planets' transits is fairly large (0.4\% and 1.7\%). This has allowed
us to detect both planets' transits from ground-based photometry,
refining their radii and orbital ephemerides. In the future, radial
velocity observations and transmission spectroscopy can both probe
these planets' bulk interior and atmospheric compositions, and
additional photometric monitoring would be sensitive to even smaller
transiting planets.
\end{abstract}


\section{Introduction}
\label{sec:intro}

Cool, low-mass stars --- M dwarfs --- are more numerous and host
more short-period planets per star than the more massive stars that
host most of the known planets
\citep{bonfils:2013,dressing:2015,mulders:2015b}. Whether they are
seen to transit, inferred from radial velocity spectroscopy, or
detected via gravitational microlensing, exoplanets tend to be easier
to characterize when they orbit M dwarfs instead of larger, hotter,
more massive stars. These red dwarfs are therefore popular targets for
exoplanet surveys of all types.

The {\em Kepler} mission surveyed several thousand M dwarfs for
transiting exoplanets and revealed that planet occurrence rates
increase with decreasing stellar mass and $T_\mathrm{eff}$ for
$P<1$\,yr \citep{howard:2012}.  However, most of Kepler's M dwarfs
were early-type: fewer than $600$ had $T_\mathrm{eff}<3300$~K
\citep{dressing:2013,morton:2014,hardegree-ullman:2019} and would
therefore be in the regime of stars with fully convective interiors.
Just seven stars cooler than $T_\mathrm{eff} < 3100$~K are known to
host planets\footnote{According to the NASA Exoplanet Archive, June
  2019}. Coolest of these is TRAPPIST-1 \citep[$T_\mathrm{eff} \approx
  2600$~K, $0.08M_\odot$;][]{gillon:2017}, whose seven transiting
planets hint that the number of planets per star may be high for the
very lowest-mass stars and which has sparked a flurry of theoretical
and observational follow-up studies.

It remains an outstanding question as to whether planet occurrence
continues to increase toward the lowest stellar masses (or beyond: do
brown dwarfs host planets?), and how these planets compare to those
orbiting hotter stars.  Although not a statistical mission, the {\em
  TESS} nearly-all-sky transit survey \citep{ricker:2014} can help to
answer this question. TESS will survey 70\% of the sky over its two
year prime mission, and is therefore well positioned to aid the search
for planets around nearby M dwarfs. Compared to {\em Kepler}, a much
larger percentage of the {\em TESS} project's high-priority target
list consists of M dwarfs, including some of the latest-type M
dwarfs. One of the first planets discovered by {\em TESS}, LHS~3844b
\citep{vanderspek:2019}, orbits an M5V star and is consequently
proving an excellent target for detailed characterization.

We report here the statistical validation of two exoplanets orbiting
\starname\ (2MASS J11024596-1624222, TIC 181804752), which was
recently observed by {\em TESS}. At $T_\mathrm{eff}=2960$~K this is
the third-coolest star known to host exoplanets. Thus like TRAPPIST-1
\citep{gillon:2017}, GJ~1214 \citep{charbonneau:2009}, Proxima
Centauri \citep{anglada-escude:2016}, and other similar systems, it
represents another rare laboratory to study exoplanets around the very
smallest stars.

\section{Observations}
\label{sec:obs}
Our target was first identified as a star of more than average
interest by \cite{luyten:1979}, who noted its high proper motion and
red color as part of the Luyten Palomar survey. Therefore we
henceforth refer to the star as \starname.  The star's properties were
more recently estimated in the \TESS\ Cool Dwarf Catalog
\citep{muirhead:2018}. It was found to be an attractive target for
\TESS\ transit photometry, and was scheduled for observations at a
two-minute cadence during Sector 9 of the \TESS\ prime mission on
account of its inclusion on the high-priority \TESS\ Candidate Target
List \citep{stassun:2018b} and as part of \TESS\ Guest Investigator
program GO~11180 (PI Dressing).  We note that all data products used
in the succeeding sections have been made available to the community
on the ExoFOP-TESS
website\footnote{\url{https://exofop.ipac.caltech.edu/tess/target.php?id=181804752}}.
A summary of all relevant stellar and planet properties are given in
Tables~\ref{tab:stellar} and~\ref{tab:planet}, respectively.

\subsection{{\em TESS} Transit Photometry}
\begin{figure}[b!]
\includegraphics[width = 0.92\textwidth]{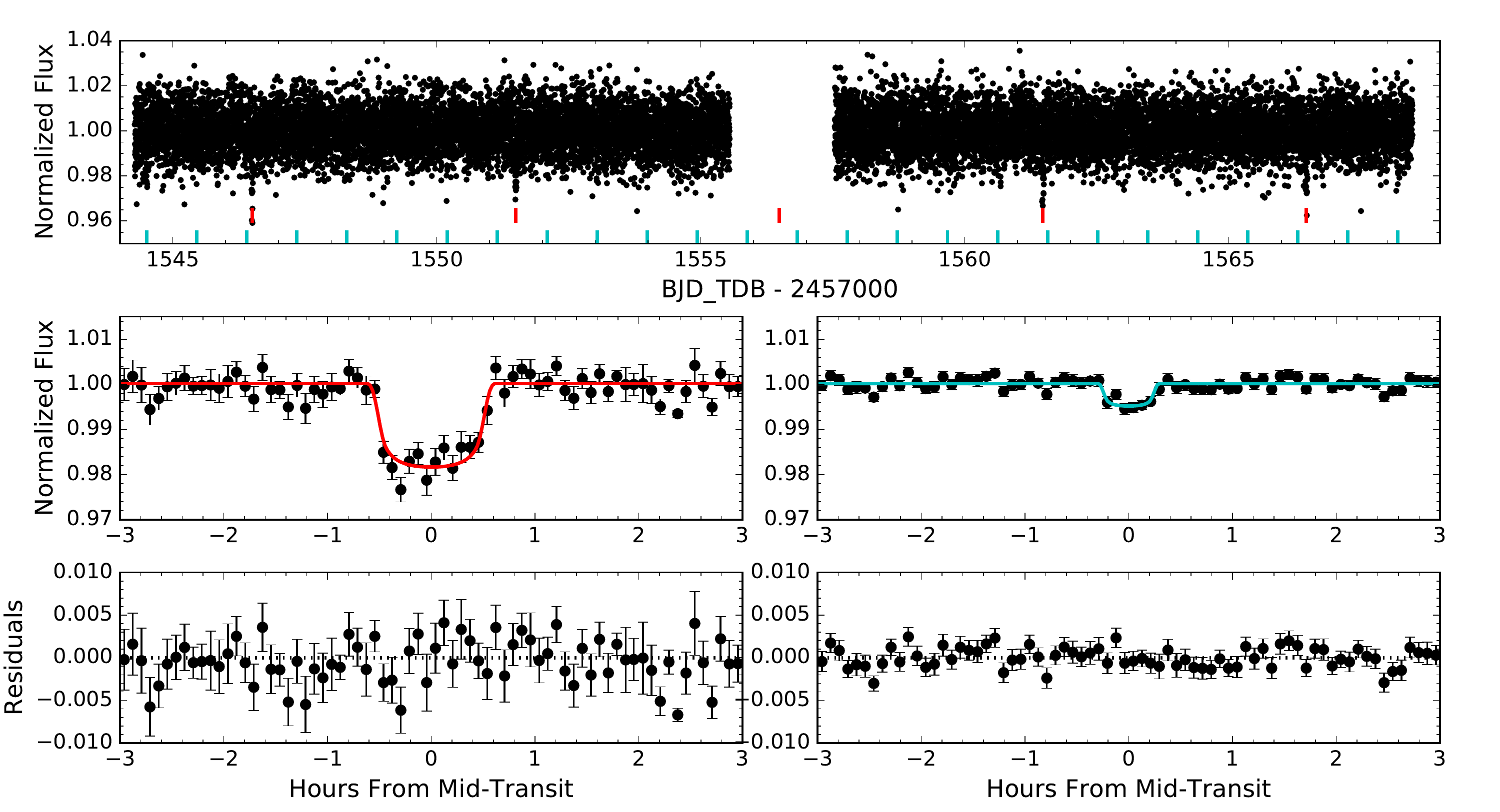}
\caption{{\em Top}: \TESS\ short-cadence photometry of
  \starname. Vertical ticks indicate the locations of each planets'
  transits.  {\em Middle}: Phase-folded photometry (binned to
  five-minute intervals, with error bars indicating the standard error
  on the mean in each bin) and best-fit light curves for each
  planet. {\em Bottom}: Residuals to the transit fits. \label{fig:tlc} }

\end{figure}

\TESS\ observed \starname\ nearly continuously from March 1--25, 2019
at a two-minute (``short'') cadence. Initial data processing was
similar to that of $\pi$~Men \citep{huang:2018}, LHS~3844b
\citep{vanderspek:2019}, and other recent \TESS\ discoveries. Analysis
by the \TESS\ Science Processing Operations Center
\citep{jenkins:2002,jenkins:2016} identified two possible planetary
signals, and human vetting of the data reports \citep[N.\ Guerrero et
  al., in prep.,][]{twicken:2018,li:2019} resulted in the announcement
of planet candidates TOI-736.01 and~.02. The \TESS\ PDC\_SAP light
curve \citep{smith:2012pdc,stumpe:2014} is shown in
Fig.~\ref{fig:tlc}. Individual transits of the larger, longer-period
TOI-736.01 are visible by eye, while the shallower, shorter-period TOI-736.02
can only be seen in the phase-folded photometry. We also used the
\TESS\ Quick-Look Pipeline (Huang et al., in prep.) to confirm that
the transit-like events are visible in the \TESS\ long-cadence data,
but the short transit durations make those data unsuitable for a
detailed light-curve analysis.

We used the short-cadence light curve to conduct a transit analysis of
both signals, using the same software as described by
\cite{crossfield:2016,crossfield:2017}. The only difference from those
analyses of long-cadence K2 photometry is that we now numerically
integrate each model light curve over just the two-minute (not
30-minute) duration of each point. As in those analyses, we impose
priors on the quadratic limb-darkening coefficients. Based on the
stellar parameters derived below and the distribution of coefficients
from \cite{claret:2018}, we adopted Gaussian priors of $u_1=0.26 \pm
0.06$ and $u_2=0.55 \pm 0.07$.  Our best-fit transit models for both
signals are shown in Fig.~\ref{fig:tlc}, and the model parameters are
listed in Table~\ref{tab:planet}.  The best-fit mid-transit times (in
BJD$_\mathrm{TDB}$) from the \TESS\ photometry are
$T_0=2458546.50885\pm0.00096$ and $2458543.5584 \pm 0.0017$ for
TOI-736.01 and~.02, respectively.

We performed an independent check on our light curve analysis, using
an approach similar to that described by \cite{chontos:2019}. This
parameterization fits for $P$, $T_0$, quadratic limb-darkening
coefficients ($u_1,u_2$), $\rho_{*,circ}$, $b$, $R_P/R_*$, and a
photometric normalization.  Again, we assume a linear ephemeris,
circular orbit, and quadratic limb darkening law with Gaussian priors
imposed.  Additional priors were used to constrain $u_1$ to the
interval [0,2], $u_2$ to [-1,1], and to ensure $\rho_{*,circ} > 0$. We
explored the parameter space using the \texttt{emcee} Markov chain
Monte Carlo algorithm \citep{foreman-mackey:2012}, initializing 100
walkers and having  each take 20000 steps. A burn-in phase of 5000 steps was
removed before compiling the final posterior distribution for each
parameter. Our two transit analyses are consistent,  agreeing to
within 1$\sigma$ for all of the derived quantities. 

Following the methodology of \cite{berardo:2019}, we also conducted a
search for transit timing variations (TTVs) by fitting each transit of
TOI-736.01 individually, allowing only the transit midpoints to vary. By
comparing these individual times to a linear ephemeris, we
conclude that there are no significant TTVs for either of the signals.

\subsection{Stellar Properties From Archival Photometry}



We estimate the spectral type of \starname\ by comparing the Gaia DR2
photometry and parallax \citep{gaia:2018} to color-type and absolute
magnitude-type relations \citep{kiman:2019}. The six possible
relations indicate an M dwarf with subclass mean and standard
deviation of $6.1 \pm 0.7$, which we adopt as the spectral type of our
target. \starname\ is not elevated above the main sequence when
plotted on an optical-infrared color-magnitude diagram, indicating
that the star is not an unresolved near-equal mass binary --- unless
it has a markedly sub-solar metallicity, which we rule out below.

We estimate the stellar mass using the $K_S$-band mass-luminosity
relation of \cite{mann:2019}. The statistical and systematic
uncertainties on the derived mass are both 0.0033~$M_\odot$, so we
report $M_*=0.139 \pm 0.005$. This mass is consistent with that
derived from earlier $V$- and $K$-band mass-luminosity relations
\citep{benedict:2016}.

We estimate the stellar radius using the absolute-magnitude
vs.\ radius relations of \cite{mann:2015}. Those authors indicate that
the $JHK_S$ relations are their most precise.
Taking the weighted mean of the three derived radii, we
find $R_* = 0.171 \pm 0.018$. This radius and the mass are consistent
with the mass-radius relation for low-mass stars \citep{mann:2015}.

We estimate the stellar effective temperature using calibrated
photometric color relations. \cite{mann:2015} demonstrate a tight
correlation between $T_\mathrm{eff}$ and $(V-J)$, $(r-z)$, and
$(r-J)$. These relations all have intrinsic scatters of about 55~K, so
we take the mean of the three derived temperatures to find
$T_\mathrm{eff} = 2960 \pm 55$~K. This value is  consistent
with temperatures estimated from tabulated photometric relations and from
our derived spectral type \citep{kraus:2007,pecaut:2013}.

We estimate the stellar metallicity using photometric relations. The
approach of \cite{schlaufman:2010} gives the best agreement with
near-infrared spectroscopic metallicities \citep{rojas-ayala:2012}. Using
their methodology, we find $\Delta(V-K_S)=0.2$, which implies
[Fe/H]\,$=-0.02 \pm 0.21$ (accounting for the relation's intrinsic
scatter and our uncertainty on $V$). Comparison to the $(G_R - J),
M_K$ color-magnitude diagram of \cite{kesseli:2019} indicates a
consistent metallicity of $-0.5 \pm 0.5$. We report the weighted mean
of these two independent estimates, [Fe/H]\,$=-0.09 \pm 0.19$.

As previously noted, \starname's properties were also estimated by
\cite{muirhead:2018} using broadband photometry but without the
benefit of Gaia DR2.  All of their stellar parameters are within
1$\sigma$ of ours (Table~\ref{tab:stellar}).


\subsection{High-Resolution Spectroscopy}
To further characterize the system and check for any evidence of
spectroscopic binaries that could indicate a non-planetary origin for
the transit signals, we obtained high-resolution spectra from
Keck/HIRES and Subaru/IRD.

\subsubsection{Keck/HIRES}
We acquired an optical  spectrum using Keck/HIRES \citep{vogt:1994} on
2019/06/12. The observation took place in $1.0\arcsec$ effective
seeing and using the C2 decker without the iodine gas cell, giving an
effective resolution of $\lambda/\Delta \lambda \approx
55,000$. We exposed for 1386~s and obtained S/N of roughly 30 per
pixel. Data reduction followed the standard approach of the California
Planet Search consortium \citep{howard:2010b}.

Using the approach of \cite{kolbl:2015}, we examined our spectrum for
secondary components that would indicate the presence of another
star. We found no evidence of additional lines down to the method's
standard sensitivity limit of $\Delta V=5$~mag for $\Delta v >
10$~km~s$^{-1}$, consistent with \starname\ being a single, isolated
star.  Finally, we measured \starname 's absolute radial velocity
following \cite{chubak:2012}, finding $14.1 \pm 0.1$~km~s$^{-1}$.

\begin{figure}[b!]
\includegraphics[width = 0.9\textwidth]{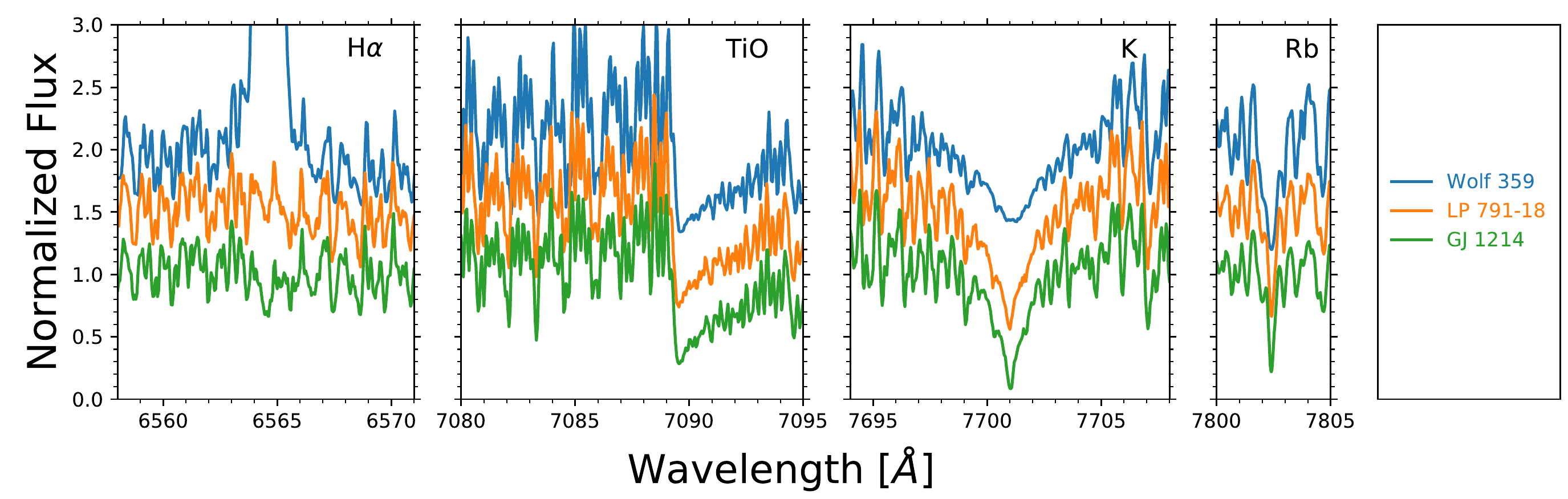}
\caption{Comparison of Keck/HIRES spectra of \starname\ (orange) with
  GJ~1214 (green) and Wolf 359 (blue) in the vicinity of the expected
  locations of H$\alpha$, TiO bands, K~I (7701.0\AA), and
  Rb~I (7802.4\AA).  No secondary spectral lines are detected.
  \label{fig:hires}}
\end{figure}

We compare our HIRES spectrum with several archival spectra of other cool M dwarfs (Fig.~\ref{fig:hires}).
\starname's spectrum is very similar to an archival HIRES spectrum of
the M5V GJ~1214 (suggesting generally similar metallicity and
temperature), but with slightly broader K~I line (consistent with
higher surface gravity). Aside from this pressure-broadened line, the
widths of weaker lines in these two stars are indistinguishable,
consistent with a low projected rotational velocity for both stars
\citep[$v \sin i < 2$~km~s$^{-1}$ for
  GJ~1214;][]{charbonneau:2009}. In comparison, \starname's lines are
noticeably narrower than those seen in archival HIRES spectra of the
M6V Wolf~359, which has $v \sin i < 3$~km~s$^{-1}$
\citep{jenkins:2009}. However, we note that if \starname\ is typical
of old, low-mass M dwarfs and has a rotation period of roughly
100~days \citep{newton:2017}, then $v \sin i \sim 0.1$~km~s$^{-1}$.
The lack of H$\alpha$ emission (see Fig.~\ref{fig:hires}) in a star of
this mass also indicates that the star likely has a rotation period $>
100$~days \citep{newton:2017}. Thus the value reported in
Table~\ref{tab:stellar} of $v \sin i <2$~km~s$^{-1}$ should be taken
as a conservative upper limit.

Several additional lines of evidence are consistent with this
interpretation of \starname\ as a relatively old star. First, we do
not detect the Li line, indicating that \starname\ is older than
0.5~Gyr \citep{reiners:2009}. Second, H$\alpha$ is not seen in
emission (see Fig.~\ref{fig:hires}); though there is no true continuum
against which to compare the line, H$\alpha$ appears to be slightly in
absorption (with a similar depth and profile to GJ~1214), suggesting a
long rotation period (as noted above) and therefore an age of several
Gyr \citep{newton:2016}.
Furthermore, comparison of our inferred $M_*$, $R_*$, and $L_*$ to
evolutionary models of ultracool dwarfs \citep{fernandez:2019} also
indicates an age $\gtrsim 0.4$~Gyr. In addition, the Galactic space
velocity of \starname\ is not consistent with any of the known young
moving groups or associations \citep{gagne:2018}, with a 95.7\%
likelihood of it being a field star \citep[according to
  \texttt{BANYAN-SIGMA};][]{gagne:2018}; its dynamics are consistent
with the thin disk rather than with the thick disk or halo.  All these
points, combined with the lack of large-amplitude variations in
\TESS\ or MEarth photometry \citep{newton:2018} suggest that
\starname\ has an age $\gtrsim0.5$~Gyr, and likely at least several
Gyr.

\subsubsection{Subaru/IRD}
We observed \starname\ with the InfraRed Doppler instrument
\citep[IRD;][]{kotani:2018} behind an adaptive optics system
\citep[AO188;][]{hayano:2010} on the Subaru 8.2 m telescope on
2019/06/17. 
We took 3 spectra with exposure times of 600~s each, with airmass of
1.8--2.0, covering the wavelength range from 0.95--1.76\,$\mu$m at
spectral resolution $\approx70,000$.  We processed the spectra using
standard tools.
An absolute wavelength solution was assigned using Th-Ar calibration
spectra \citep{kerber:2008} and laser frequency comb spectra, both of
which were previously taken during daytime observations.

We combined our three exposures into one template spectrum for visual
inspection of possible contamination of additional faint star.
Achieved signal-to-noise ratio around the peaks of blaze function in
the combined spectrum is roughly 30 in $Y$, 50 in $J$, and 80 in $H$.
As with our HIRES analysis, we do not see any evidence of
contaminating lines in the spectrum.

\subsection{High-resolution Imaging}
\TESS\ has large pixels ($21\arcsec$ across), which are large enough
to contain many additional stars that could potentially be the source
of the detected transits.  To identify any additional stars around
\starname, we obtained several sets of high-resolution imaging data,
as described below.

\subsubsection{Gemini/`Alopeke Optical Speckle Imaging}
\begin{figure}[b!]
\includegraphics[width = 0.9\textwidth]{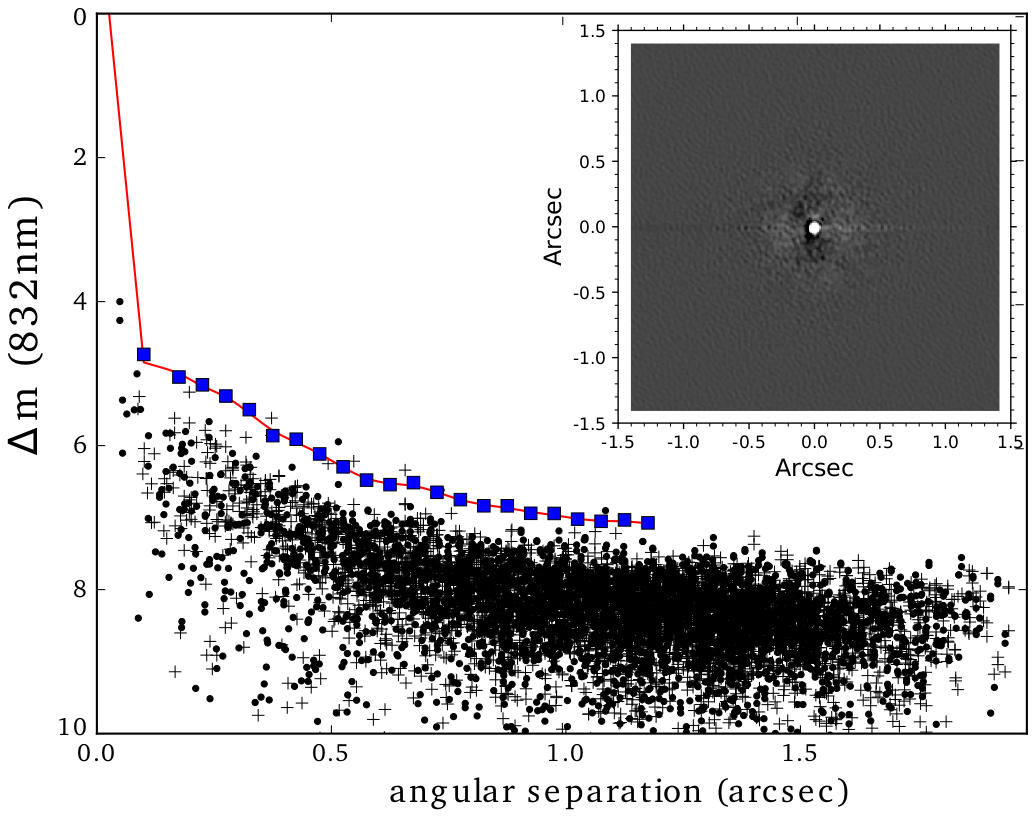}
\caption{Red-optical contrast limits set by our Gemini/`Alopeke
  speckle imaging data, with the reconstructed image shown at inset. No
  secondary sources are detected.
  \label{fig:speckle}}
\end{figure}

We observed \starname\ with the `Alopeke speckle imaging instrument
\citep{howell:2011,scott:2018} on the Gemini-North 8.1 m telescope on
2019/06/08. The observations consisted of 18 simultaneous image sets
of one thousand 60~ms frames in narrow band filters centered at 562
and 832~nm in good observing conditions, with the native seeing
measured to be $0.4\arcsec$. Because our target is quite red, the data
at 832~nm are superior to those at 562~nm.  The speckle images were
reduced alongside the point source calibrator star HR~4284 standard
reduction procedures \citep{howell:2011,matson:2018}.  Data products
include the power spectrum of the speckle patterns of
\starname\ divided by those of HR~4284, and a reconstructed image of
the $2.5\times2.5\arcsec$ field centered on the target (shown in
Fig.~\ref{fig:speckle}).

These data products were inspected for neighboring sources and none
were found.  Contrast curves were produced from the reconstructed
images by normalizing the peak flux of the star and determining the
standard deviation in magnitudes among local minima and maxima in the
background noise as a function of angular separation from the star.  A
flux level 5$\sigma$ brighter than the mean of the local extrema is
used to define the limiting contrast relative to the \starname. At
832~nm we achieve a contrast of 4.8~mag at $0.1\arcsec$, increasing
steadily to 7.0~mag at $1.2\arcsec$, as shown in
Fig.~\ref{fig:speckle}.

\subsubsection{Keck/NIRC2 AO Imaging and Aperture Masking}
On 2019/06/12 we obtained laser guide star adaptive optics (LGS-AO)
imaging \citep{wizinowich:2000} and non-redundant aperture masking
interferometry \citep[NRM;][]{tuthill:2006} of \starname\ in two
visits separated by 15 minutes with Keck/NIRC2. The observations were
taken in vertical angle mode without dithering and using the K'
filter, and the NRM used the 9-hole mask.  We also observed two nearby
calibrator stars. In all cases we used the smallest pixel scale of
9.952 mas~pix$^{-1}$. For imaging we took 12 exposures, each with 20
coadds of 1\,s duration and four Fowler samples. For NRM we took 16
interferograms, each with one coadd lasting 20\,s and comprising 64
Fowler samples. For the first calibrator we obtained eight images and
eight interferograms in similar setups. For the second calibrator we
obtained eight images and four interferograms.

On 2019/06/13 we obtained additional LGS-AO imaging of \starname\ and
the first calibrator star, again in K' at the same pixel scale. The
observations were taken in position angle mode, rotated to align the
+y axis of NIRC2 with North. We observed in a 3-point dither pattern
that avoided the NIRC2 bad quadrant while stepping the target in
offsets of 1.0\arcsec, 1.5\arcsec, and 2.0\arcsec; we did not dither
on the calibrator.
We took  20 exposures of \starname\ using the same settings as
in the preceding night, and took eight exposures of the first
calibrator.

We reduced each frame and searched the resulting data for companions
following \cite{kraus:2016}.
We used two different strategies for PSF subtraction, applied
individually to each image. To search for faint, wide companions at
$>$500 mas, we subtracted a model constructed from the
azimuthally-averaged flux profile of \starname. This added no
additional noise at wide separations, but left the speckles in place,
making it non-ideal for detecting close-in companions. To probe
smaller inner working angles we then also subtracted a scaled version
of the best-fitting empirical PSF taken from the set of all imaging
observations of the calibrator stars. We measured the flux as a
function of position within each residual image using 40~mas (radius)
apertures centered on every image pixel, and stacked the
Strehl-weighted significance maps of each frame in order to compute
the final significance map for potential detections around
\starname. We measured our detection limits from the distribution of
confidence levels among all apertures in a series of 5-pixel annuli
around the primary.  No apertures contain a statistically significant
excess of flux within the NIRC2 field of view, and hence there are no
detected astrophysical sources. We pursued similar analysis for both
calibrators and found that they also have no astrophysical sources
within the observed FOV.

The non-redundant masking observations use a pupil plane mask to
resample the telescope into a sparse interferometric array. This
allows the use of the complex triple product, or closure-phase
observable, to remove non-common path errors produced by atmospheric
conditions or variable optical aberrations. To remove systematics in
the closure-phase observable, the observation of \starname\ was paired
with  observations of the two calibrator stars, both of
which have similar color and brightness and are located within $1
\degr$ of \starname. Our analysis followed the methods described in
the appendix of \cite{kraus:2008}.  Binary-source models were fit to
the calibrated closure phases to search for significant evidence of
binarity, and the detection limits were calibrated by repeatedly
scrambling the phase errors and determining the distribution of binary
fits. Again, no sources were detected in the masking data for
\starname.

\begin{figure}[b!]
\includegraphics[width = 0.9\textwidth]{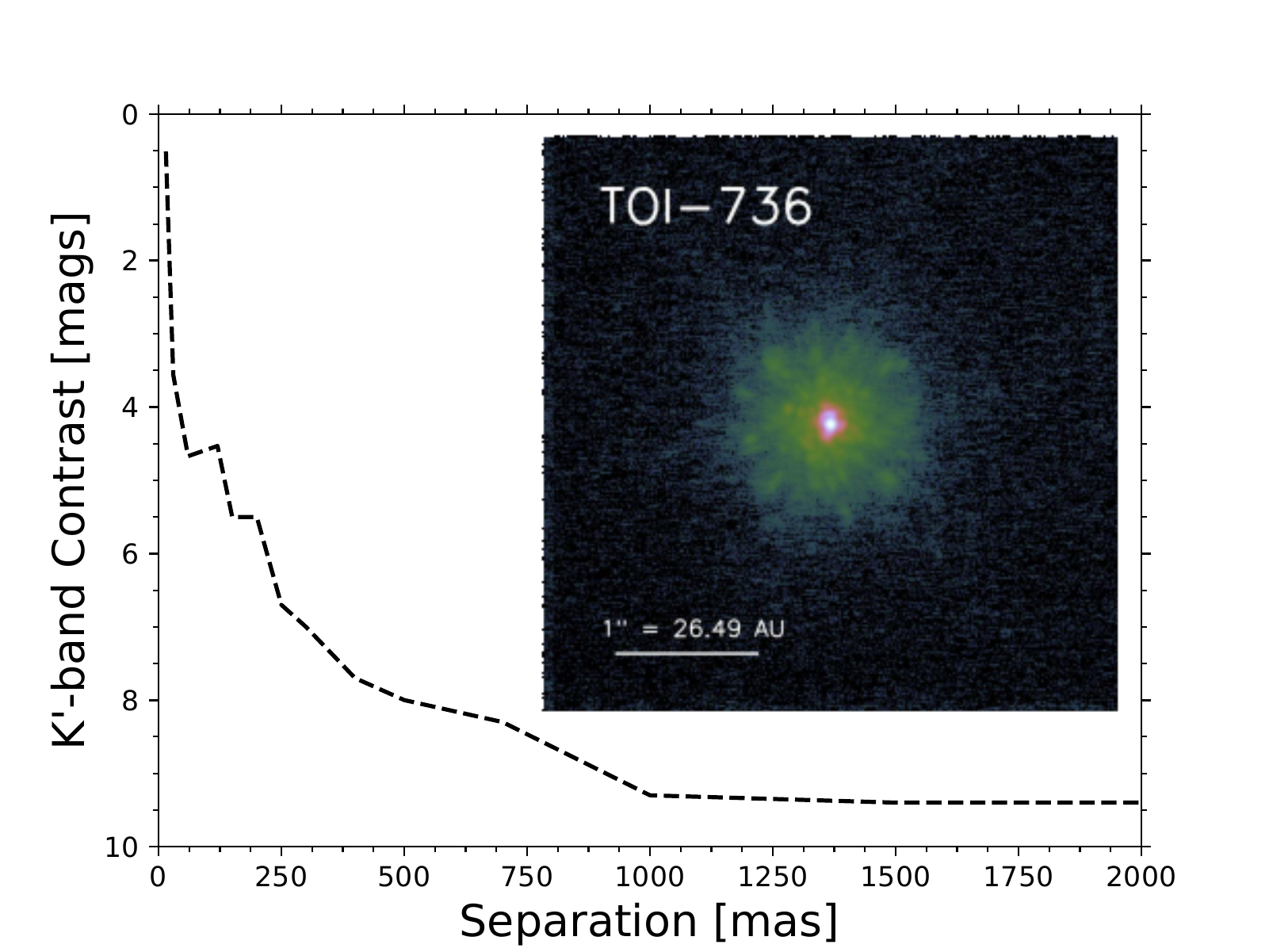}
\caption{K$^\prime$-band contrast limits set by our Keck/NIRC2 adaptive
  optics imaging and aperture masking data, with  the
  imaging data shown at inset. No secondary sources are detected.
  \label{fig:nirc2}}
\end{figure}

Fig.~\ref{fig:nirc2} shows the effective contrast achieved by  our NIRC2
observations.  The combination of aperture masking and
imaging data excludes many companions to \starname, reaching
contrast ratios of $\Delta K' = 3.56$ mag at $\rho = 20$ mas, $\Delta
K' = 4.67$ mag at $\rho = 40$ mas, $\Delta K' = 5.5$ mag at $\rho =
150$ mas, $\Delta K' = 6.6$ mag at $\rho = 200$ mas, and an ultimate
limiting magnitude of $\Delta K' = 9.3$ mag at $\rho > 1\arcsec$.
Comparison to the MIST isochrones \citep{dotter:2016,morton:2015b} for
all stars at the same distance and with $t>100$~Myr shows that our
suite of high-resolution imaging data rule out all companions down to
the H-burning limit from the NRM inner limit of 20~mas (0.8 AU) out to
the edge of the NIRC2 FOV in the dithered data set ($9\arcsec$,
230~AU).  We rule out all companions with spectral types $>$L5 beyond
1.1~AU, $>$T4 beyond 5.4~AU, and $>$T8 beyond 22~AU
\citep{dupuy:2012,dupuy:2017}.

\section{Ground-based transit photometry }
\subsection{Las Cumbres Observatory}
We also observed both transit signals using 1.0\,m telescopes of the Las
Cumbres Observatory \citep[LCO;][]{brown:2013}.  We used the {\tt TESS
  Transit Finder}, a customized version of the {\tt Tapir} software
package \citep{Jensen:2013}, to schedule the photometric follow-up
observations. All observations used a $4096^2$ LCO SINISTRO camera
with an image scale of 0$\farcs$389 pixel$^{-1}$ resulting in a
$26\arcmin\times26\arcmin$ field of view.

We acquired one transit light curve of TOI-736.01 on 2019/06/16 at the
South Africa Astronomical Observatory, and two light curves of
TOI-736.02 on 2019/06/11 from two telescopes at Siding Spring.  The
transit of TOI-736.01 comprised 114 images in Bessel-I band using
60\,s exposures, for a total duration of 169 minutes.  The two
transits of TOI-736.02 included 99~min in $I_C$ band with 60\,s
exposures, and 192~min in Sloan $i^{\prime}$ with 100\,s
exposures. The target star had an average FWHM of $2.2\arcsec$,
$2.4\arcsec$, and $1.4\arcsec$, respectively. The nearest known Gaia
DR2 star is $15\arcsec$ from \starname: it has $\Delta
G_\mathrm{RP}=4.3$ and so is too faint to significantly dilute the
\TESS\ transit photometry, (and our high-resolution imaging detected
no additional companions), so the LCO follow-up apertures are negligibly
contaminated by neighboring stars.

\begin{figure*}[b!]
\includegraphics[width = 0.5\textwidth]{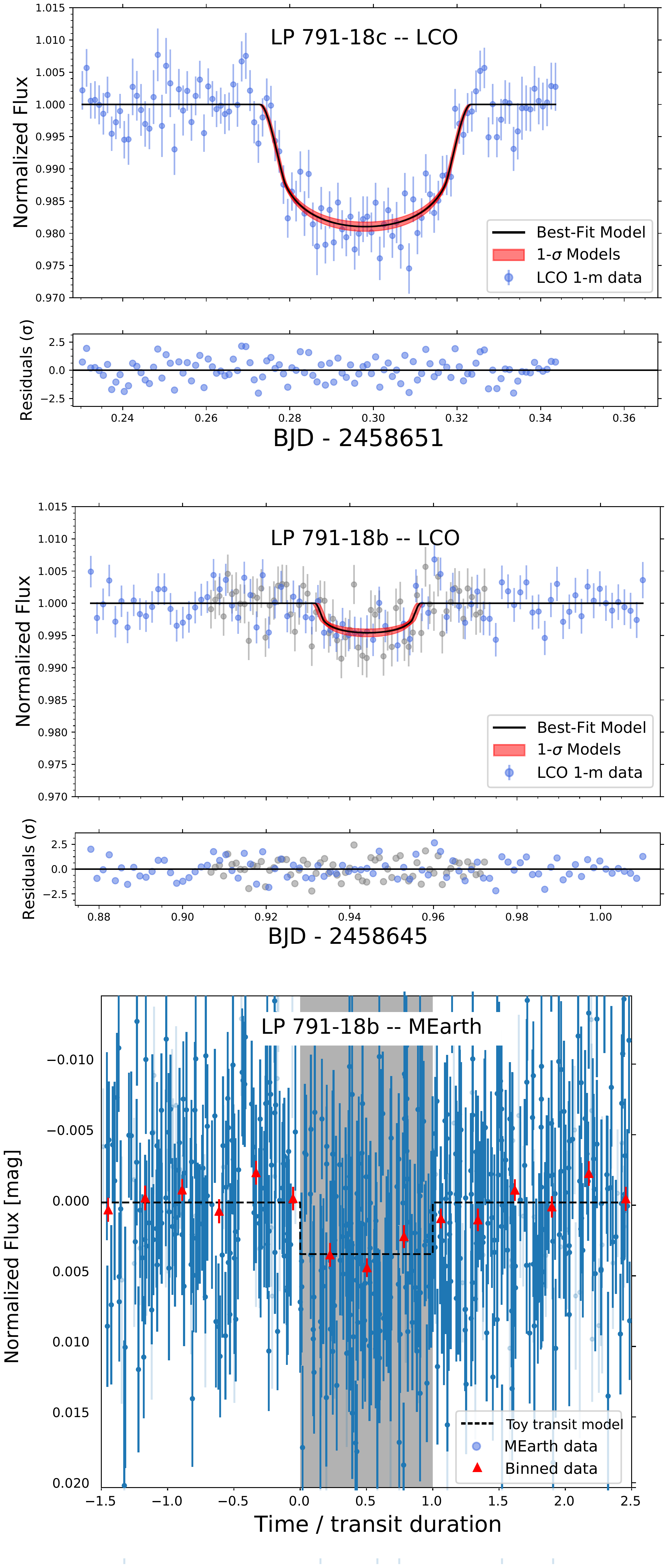}
\caption{Ground-based photometry of \starname. {\em Top:} one LCO 1\,m
  transit of \starname c. {\em Middle:} two LCO 1\,m transits of
  \starname b.  {\em Bottom:} Phase-folded MEarth-South photometry of
  \starname b during transit.  \label{fig:lco}}
\end{figure*}

All data were calibrated by LCO's standard BANZAI pipeline and the
photometric data were extracted using {\tt AstroImageJ}
\citep{Collins:2017}. In all cases, the target star light curve shows
a clear transit detection, while a search for eclipsing binaries
within $2.5^\prime$ that could have caused the transit signal reveals
nothing.  The transit signal can be reliably detected with apertures
having radii as small as $1\farcs95$, but systematic effects start to
dominate for smaller apertures.  Fig.~\ref{fig:lco} shows our LCO
photometry, in which transits are clearly visible.

We model all three LCO light curves with \texttt{BATMAN}
\citep{kreidberg:2015b} keeping all parameters --- except for $R_p
/R_*$ and the mid-transit time --- fixed to the values derived from
the \TESS\ light curve (Table~\ref{tab:planet}). We also include a
linear airmass correction model to account for the out-of-transit
baseline, and limb darkening was calculated using \texttt{LDTK}
\citep{parviainen:2015} based on the parameters in
Table~\ref{tab:stellar}. For TOI-761.01 we measure $R_p/R_*= 0.1233\pm
0.0024$ and $T_0 = 2458651.29807 \pm 0.00041$, while for TOI-761.02 we
measure $R_p/R_* = 0.0624 \pm 0.0044$ and $T_0 = 2458645.94429\pm
0.00078$ (times in $BJD_\mathrm{TDB}$).   The LCO transit depths
are all consistent with those measured by
\TESS.

\subsection{MEarth-South}
\starname\ is also a target of the MEarth transit survey
\citep{nutzman:2008,irwin:2015}. The MEarth data set consists of 4534
photometric observations obtained with the MEarth-South
telescope array between May 2015 and June 2019. These photometric data
do not reveal any coherent periodic variations that would indicate a stellar rotation period.

A Box-Least Squares (BLS) search of the MEarth photometry independently reveals a signal with a
period of 0.948002~days (shown in Fig.~\ref{fig:lco}), consistent with
the TESS ephemeris of TOI-736.02. Due to the near-integer orbital
period of TOI-736.01, its transits cannot be recovered in the MEarth
photometry (though some combinations of $P$ and $T_0$ are ruled out).

We used \texttt{BATMAN}  and \texttt{emcee}
 to model the transits of TOI-736.02 using
the combined TESS and MEarth photometry. The values of $a/R_*$ and $i$
were kept fixed at the values obtained from the TESS analysis
(Table~\ref{tab:planet}) while $T_0$, $P$, and $R_p/R_*$ were allowed
to vary. From this analysis we measure $T_0 = 2458645.9434 \pm
0.0013$, $P=0.9480048 \pm 0.0000058$~days, and $R_P/R_* =
0.059^{+0.0033}_{-0.0042}$.

\subsection{Refined transit parameters}
We use the LCO and MEarth-South data sets to improve the ephemerides
and transit depths of both transit signals by taking the weighted mean
of $R_P/R_*$ and combining the $T_0$ values using weighted
least-squares and assuming a linear ephemeris.  For TOI-736.01
(undetected by MEarth) we use the results of the \TESS\ and LCO
transit analyses, while for TOI-736.02 we use the results of the LCO
transit and the combined {\em TESS}+MEarth analysis.

We find that for each signal, $R_P/R_*$ is consistent across all our
analyses and $T_0$ is consistent with a linear ephemeris.  Including
the ground-based data decreases the uncertainty on $P$ by an order of
magnitude (for TOI-736.01) and two orders of magnitude (for
TOI-736.02). We report the final values of $R_P/R_*$, $R_P$, $T_0$,
and $P$ for both TOIs in Table~\ref{tab:planet}.


\section{Statistical Validation of the Candidates}
Although transits are clearly seen by \TESS\ (Fig.~\ref{fig:tlc}) and
from the ground (Fig.~\ref{fig:lco}), many \TESS\ candidates have been
identified as false
positives\footnote{\url{https://exofop.ipac.caltech.edu/tess/}} and so
we must verify that the observed signals are planetary in origin.
Since the precise Doppler spectroscopy needed to confirm these signals
as planets will likely need to wait until \starname\ rises again for the next
season, we  demonstrate below that the signals are far
more likely to be planetary than of any other origin.  Below, we
consider whether \starname\ could be blended with a background
eclipsing binary, and then whether \starname\ itself could be a
multiple star system.  We find that both scenarios are unlikely,
indicating that our planet candidates are likely to be true transiting
planets.

\subsection{Independent Signal Validation}
We used the Discovery and Validation of Exoplanets
tool\footnote{\url{https://github.com/barentsen/dave}}
\cite[\texttt{dave};][]{kostov:2019a,kostov:2019b}, along with the
short-cadence pixel files and photometry, to independently estimate
the quality of the candidate planet signals. We find no significant
secondary eclipses or odd-even differences (which would otherwise
indicate an eclipsing binary instead of a transiting planet) for
either TOI. We find no significant photocenter shift (which would
indicate a blend of multiple stars, and possible source confusion) for
TOI-736.01.  For TOI-736.02 the individual difference images per
transit are too noisy for \texttt{DAVE} to provide an accurate
photocenter analysis. Nonetheless, neither candidate shows indications
of being a false positive.

The \texttt{dave} results are consistent with the \TESS\ project's
data validation tests \citep{twicken:2018,li:2019}, which both TOIs
passed. These tests include the odd-even transit depth test, the weak
secondary test, the ghost diagnostic test, the difference image
centroid offset test (0.35 and 0.5 sigma for the TIC offset for
candidates 1 and 2, respectively, representing less than $1\arcsec$
offsets from the TIC position), and the statistical bootstrap test
(which gave $7 \times 10^{-73}$ and $3 \times 10^{-15}$ for TOIs 736.01
and .02, respectively).

\subsection{Unassociated background scenarios:}
Our ground-based photometry demonstrates that transits occur close to
\starname, but a background system could lie near the star and mimic
planetary transits. Given its high proper motion, \starname\ has moved
considerably since its detection by the Palomar Optical Sky Survey in
the 1950s. No source is visible in the digital POSS-I images at the
star's location during the \TESS\ epoch. By comparing these images to
SDSS DR9 photometry, we confidently exclude any background object down
to a limit of approximately $r=19.5$, $i=18.6$ (AB mags),
corresponding to $R\approx19.2$, $I_C\approx 18.3$ (Vega
mags\footnote{\url{http://www.sdss3.org/dr8/algorithms/sdssUBVRITransform.php}}).
Since the broad \TESS\ bandpass is quite red \citep{ricker:2014} we
assume that the limiting \TESS\ magnitude is $T\approx I_C \approx
18.3$, 4.8~mag fainter than \starname.  We therefore exclude all
background sources as the source of TOI-736.01's transits.

From POSS-I and {\em TESS} alone we cannot rule out all such
background scenarios for the shallower TOI-736.02 (since its transit
depth is $<10^{-0.4 \times 4.8}$), but our LCO transit observations
demonstrate that these events occur within a few arcsec of
\starname. Since our high-resolution imaging shows no additional
sources, this all but eliminates the chance that this shallower
candidate is a background system.  A 4.8~mag-fainter source could
reproduce the 736.02 transits (with depth roughly 0.4\%) if it had a
40\% (intrinsic) transit depth -- or if it had $T\approx 19.4$~mag and
were completely eclipsed.  The only allowable brightness of a
background source is $T$ in the range 18.3--19.4~mag.  We used the
TRILEGAL Galactic stellar population
simulator\footnote{\url{http://stev.oapd.inaf.it/cgi-bin/trilegal}}
\citep{girardi:2005} to find a 0.2\% chance that our LCO photometric
aperture would contain a star with this brightness. If these simulated
stars were actually equal-mass binaries with $P=0.95$~day, the average
transit probability of the ensemble (assuming $e=0$) is 17\%. The
median star in this distribution is a 0.4$M_\odot$ M dwarf, and the
tight binary fraction of such stars is about 3\%
\citep{blake:2010,clark:2012}. The product of these factors is the
likelihood that TOI-736.02 is a background false positive: this is
$10^{-5}$, so we conclude that both transit signals are unlikely to
arise from blends with a background eclipsing binary.


\subsection{Bound, Multi-star scenario:}
\label{sec:bound}
We now consider the scenario that \starname\ is itself a multiple
system with transits occurring around just one component --- this, too,
turns out to be unlikely.  M dwarfs in the Solar neighborhood with
0.075--0.3$M_\odot$ have a multiplicity fraction of about 20\%
\citep{winters:2019}. Following the parameters given in that work, we
simulated a distribution of binary companions to \starname\ with a
log-normal distribution in $a$ that peaks at 10~AU, with
$\sigma_{\mathrm{log_{10}} (a/\mathrm{AU})} = 1$, and with a
linearly-increasing mass fraction distribution from 0.1 to unity.

\begin{figure}[b!]
\includegraphics[width = 0.9\textwidth]{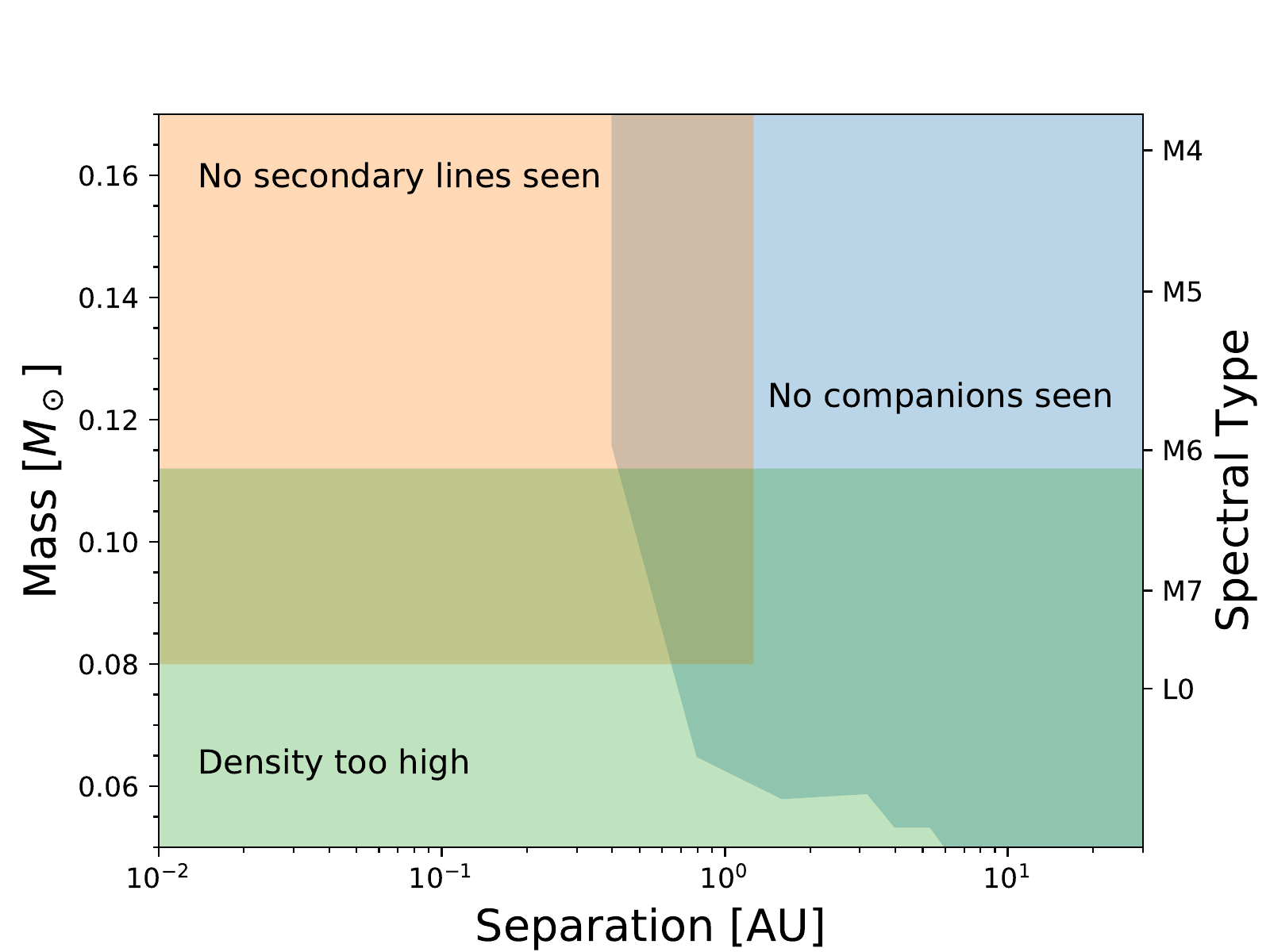}
\caption{Our analysis rules out bound companions of all types as the
  hosts of the detected transits. Clockwise from upper left: we see no
  secondary lines in our high-resolution spectra, ruling out bright,
  short-period companions; we see no companions in our high-resolution
  imaging data, ruling out long-period companions; and our transit
  analysis indicates a density than excludes objects with $<0.11
  M_\odot$. See Sec.~\ref{sec:bound} for details.
  \label{fig:companions}}
\end{figure}

We then compare this population of plausible companions to our
observations: we see no companions in our high-resolution imaging; the
system is not overluminous relative to the M dwarf H-R diagram; the
host star's density is constrained by the transit light curve analysis
\cite[$\rho_{*,circ}$ in Table~\ref{tab:planet};][]{seager:2003};
companions of later type than roughly L5 ($M\lesssim 0.03 M_\odot$)
would be too faint to be the source of the transit signals
\citep{dahn:2002}; and we see no evidence for secondary lines in our
high-resolution optical spectrum.  Fig.~\ref{fig:companions} shows
that our observations cover all relevant regions of false positive
parameters space.

After accounting for the possibility that, by chance, some wide
companions could have a very low projected separation from
\starname\ and some short-period companions could have had zero
velocity offset from \starname, we still find just a 0.7\% chance that
an additional companion is the source of the transit signals and went
unnoticed by our observations. The remaining possible configurations
involve a 0.03--0.04$M_\odot$ brown dwarf orbiting \starname\ with
$a\approx 0.7$~AU and nearly or fully eclipsed by a giant planet or
brown dwarf.

Taking the distribution of mass fractions and semimajor axes of
low-mass stellar binaries \citep{winters:2019}, and accounting for
random orbital alignments, we calculate that a star like \starname\ has
a 0.04\% chance of being in an eclipsing binary with $P\le10$~days and
companion mass 0.03--0.04$M_\odot$. This is far less than its 66\%
chance of having a planet with with $R_P<3 R_\oplus$ on a similar
period \citep{dressing:2015}. It is therefore far more likely
that \starname\ is a single planet-hosting star than that it is a
false positive with a eclipsing brown-dwarf binary. Thus we conclude
that the signals detected by \TESS\ represent exoplanets transiting
the M6V star \starname. Henceforth we denote TOI-736.02 (the smaller,
inner planet) as \starname b and TOI-736.01 as \starname c.

\section{Discussion}

\subsection{On Multiplicity and Additional Planets}
There is evidence that the multiplicity of short-period planets is
high for stars at the latest spectral types, even though few planet
host stars are known at these coolest temperatures
(cf.\ Fig.~\ref{fig:sample}). Aside from \starname, just seven
planetary systems are known with $T_\mathrm{eff} < 3100$~K. Four of
these are multiple systems: TRAPPIST-1 \citep[seven
  planets;][]{gillon:2016,gillon:2017}, YZ~Ceti \citep[three
  planets;][]{astudillo-defru:2017,robertson:2018}, Kepler-42
\citep[three planets;][]{muirhead:2012a}, and Teegarden's Star
\citep[two planets;][]{zechmeister:2019}.  Three have just a single
known planet: GJ~1214 \citep{charbonneau:2009}, LHS~3844
\citep{vanderspek:2019}, and Proxima Centauri
\citep{anglada-escude:2016}.


To verify that additional planets could exist on stable orbits around
\starname\ with $P$ between the two transiting planets, we performed a
series of N-body dynamical simulations using the Mercury Integrator
Package \citep{chambers:1999} and following the methodology of
\cite{kane:2015}. We assessed the stability of circular orbits between
the two known planets by placing a hypothetical Earth-mass planet at
$a$ in the range 0.01--0.03~AU in steps of 0.0005~AU. Each simulation
was run for $10^5$~yr, a sufficient time span given the very short
orbital periods involved.
Our simulations show that stable orbits are possible in the range
0.011--0.0255~AU (although large planets close to low-order resonance
with \starname\ c are unlikely due to the absence of observed
TTVs).

To look for additional planets, we ran a BLS analysis of the TESS
photometry but found no significant signals. We also injected a series
of planet transit signals into the \TESS\ photometry and ran a BLS
analysis on the simulated data. For an intermediate period (e.g.,
$2.5$~days), transiting planets with $R_P \gtrsim 1.2 R_\oplus$ should
have been seen in the \TESS\ data.  For planets on longer periods
(e.g., $P=7-10$~days), $R_P \gtrsim 1.4 R_\oplus$ would have been
detected. Thus planets the size of those orbiting TRAPPIST-1 would be
unlikely to have been detected by \TESS\ around \starname. There could
easily be Earth-sized planets orbiting \starname\ that went undetected
by \TESS.  In particular, the cloud-free habitable zone for a star
like \starname\ extends from approximately $P=10-30$~days
\citep{kopparapu:2013a,kopparapu:2014}, a range only poorly sampled by
the existing \TESS\ photometry.

Additional planets could   be identified by long-duration
time-series photometry (as were sought around GJ~1214; \citeauthor{fraine:2013}
\citeyear{fraine:2013}; \citeauthor{gillon:2014}
\citeyear{gillon:2014}, and seen around TRAPPIST-1;
\citeauthor{gillon:2017,luger:2017}). However, just a few degrees of
mutual misalignment between the planets' orbits would result in any
extra planets failing to transit. Assuming circular orbits, Earth-size
planets with $P=$2.5\,days and 7\,days would need to be misaligned by
2.7$^o$ and 1.3$^o$, respectively, in order not to transit.  For
reference, the mutual misalignments of the TRAPPIST-1 planets are
$<0.4^o$ \citep{gillon:2017}, while many other ultra-short period
planets have much higher mutual inclinations of 6.7$^o$
\citep{dai:2018}.

\subsection{\starname\ b and c}
These two small planets have sizes of $1.1 R_\oplus$ and $2.3
R_\oplus$, and so straddle the radius gap at about 1.8$R_\oplus$ that
separates smaller, higher-density super-Earths from larger, more
rarified sub-Neptunes \citep{fulton:2017,fulton:2018}. It remains an
open question as to whether this radius gap (measured from FGK
systems) extends to planets orbiting M dwarfs, or to planets with this
combination of small size and low irradiation.

It seems entirely likely that the masses of these new planets can be
measured in the near future, which would help to better determine
their overall composition.  By comparison to theoretical mass-radius
relations \citep{valencia:2011} we expect the two planets to have
masses of 0.5--4$M_\oplus$ (for bulk compositions ranging from
Moon-like to Mercury-like) and 5--20$M_\oplus$ (for bulk compositions
ranging from a 50-50 water-rock mix to a 0.01\% H$_2$/He veneer on a
rocky core) for planets b and c, respectively. These compositions correspond to RV
semiamplitudes of 1--9~m~s$^{-1}$ and 7--26~m~s$^{-1}$.  A mass and
radius could distinguish between different refractory compositions of
\starname b, but the degeneracies inherent in modeling larger planets
means that RV observations can constrain, but not uniquely determine,
the bulk makeup of \starname c. The star's red color and relatively
low apparent brightness means that RV follow-up is likely to be most
productive when pursued by facilities on large ($\ge$8 m) telescopes
and/or with extended coverage into the red-optical or near-infrared.

Because of the poor constraints on the impact parameters in the light
curve, we cannot deduce much about the mutual inclination of these two
planets; measurement of the Rossiter-McLaughlin effect or Doppler
tomography observations would provide an orthogonal constraint on the
dynamical architecture.  However, these measurements are probably only
feasible if \starname\ has $v \sin i\gtrsim 2$~km~s$^{-1}$, substantially
larger than expected for a quiescent star of this type (though barely
consistent with our Keck/HIRES and Subaru/IRD spectra).
Even then, the short transit durations would make it difficult to
obtain the necessary S/N in exposures that are short enough to provide
good temporal sampling of the transit.

Atmospheric characterization of these planets is also feasible. We
simulated model transmission spectra for these planets using
\texttt{ExoTransmit} \citep{kempton:2017} and assuming planet masses
of 2$M_\oplus$ and 7$M_\oplus$, and atmospheric compositions of 100\%
H$_2$O and 100$\times$ Solar metallicity, for \starname b and~c,
respectively. Our model atmospheres assumed no clouds and chemical
equilibrium, and set the 1~bar radius equal to the transit radius
observed by \TESS. These models predict peak-to-valley transmission
signals for planets b and c of roughly 150~ppm and 500~ppm, with the
difference set largely by the two models' differing mean molecular
weights. If the planets have lower masses or lower-metallicity
atmospheres than assumed above, the desired atmospheric signals would
be even stronger.

We then used
PandExo\footnote{\url{https://github.com/natashabatalha/PandExo}}
\citep{greene:2016,batalha:2017} to simulate JWST observations of a
single transit of each planet using the NIRspec prism ($0.6-5 \mu m$)
and MIRI LRS (5-12$\mu$m) instruments modes, with a baseline of equal
time to the transit time and zero noise floor. Assuming an effective
resolution of $35$ we find that the median per-channel uncertainty on
the transit depth would be 220~ppm and 150~ppm, respectively, with the
difference set by the planets' transit durations.  For the larger,
cooler \starname c JWST should be able to identify atmospheric
features in the spectrum between $1-5 \mu m$ with just a single
transit, indicating that it could be a compelling target for
atmospheric follow up. For the smaller, warmer \starname b multiple
transits would likely be needed to probe the composition of the
planet's atmosphere (if any).

\subsection{Concluding Thoughts}
\begin{figure}[b!]
\includegraphics[width = 0.92\textwidth]{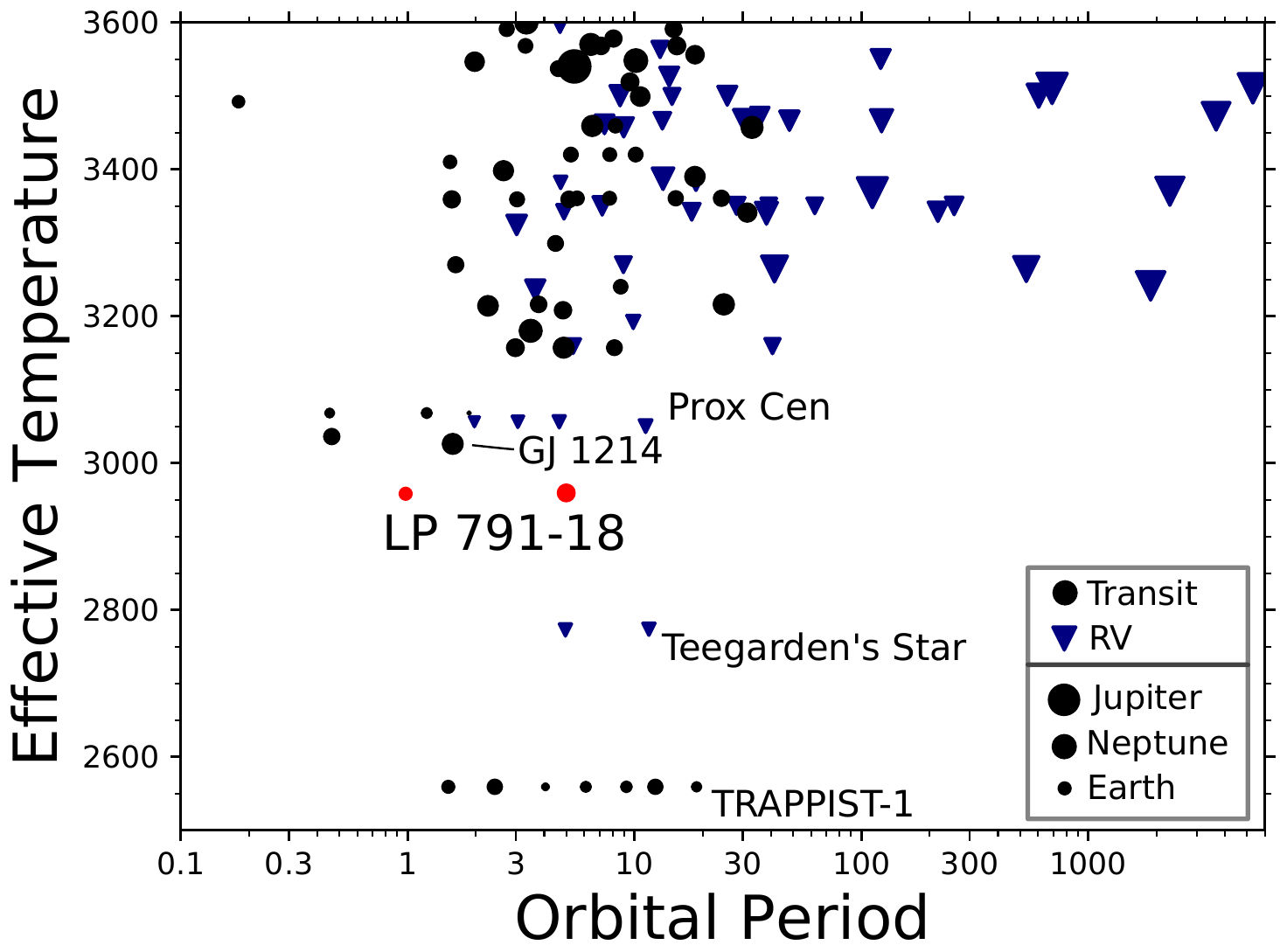}
\caption{Planets orbiting cool dwarfs. The point size increases as the
  logarithm of planet mass (inferred from radius when the mass are
  unknown). Transiting planets are shown with circles and planets not
  known to transit with triangles; our two new planets are indicated
  by red stars.\label{fig:sample}}
\end{figure}

Fig.~\ref{fig:sample} shows that \starname\ is the third-coolest star
known to host planets. The discovery of the TRAPPIST-1 system spurred
many new studies into star-planet interactions \citep{dong:2018},
multiplanet dynamics \citep{luger:2017}, atmospheric escape
\citep{wang:2018}, planet formation \citep{haworth:2018}, and
atmospheric measurements \citep{barstow:2016} of small planets around
low-mass stars.  Along with the new planets orbiting Teegarden's Star
\citep{zechmeister:2019}, \starname\ now adds another multiplanet
system against which to test these theories via the system properties
presented here, through further detailed characterization of the
planets and their host star, and by searching for additional planets
orbiting this cool dwarf.

\acknowledgments The authors thank Prof.\ N.\ Lewis for a
stimulating and thought-provoking discussion that improved the quality
of this work. We thank Hiroki Harakawa, Tomoyuki Kudo, Masashi Omiya,
Aoi Takahashi, the entire IRD team, and the Subaru IRD TESS intensive
follow-up project team for supporting Subaru IRD observation.

I.J.M.C. acknowledges support from the NSF through grant AST-1824644,
and from NASA through Caltech/JPL grant RSA-1610091.  M.R.K
acknowledges support from the NSF Graduate Research Fellowship, grant
No. DGE 1339067.  D.H. acknowledges support by the National
Aeronautics and Space Administration (80NSSC18K1585, 80NSSC19K0379)
awarded through the \TESS\ Guest Investigator Program. Work by
J.N.W. was supported by the Heising-Simons
Foundation. B.R-A. acknowledges support from FONDECYT through grant
11181295.  W.W. acknowledges support from the NSF GRFP, DGE 1650115.
N.N. is supported by JSPS KAKENHI Grant Numbers JP18H01265 and
JP18H05439, and JST PRESTO Grant Number
JPMJPR1775. A.J.C. acknowledges support from the National Science
Foundation Graduate Research Fellowship Program under Grant No. DGE
1842402. D.B. acknowledges support from an NSERC PGS-D
scholarship. The MEarth Team gratefully acknowledges funding from the
David and Lucille Packard Fellowship for Science and Engineering
(awarded to D.C.) and the National Science Foundation under grants
AST-0807690, AST-1109468, AST-1004488 (Alan T. Waterman Award), and
AST-1616624. This publication was made possible through the support of
a grant from the John Templeton Foundation. The opinions expressed in
this publication are those of the authors and do not necessarily
reflect the views of the John Templeton Foundation.

We acknowledge the use of \TESS\ Alert data, which is currently in a
beta test phase, from pipelines at the \TESS\ Science Office and at
the \TESS\ Science Processing Operations Center. This research has
made use of the Exoplanet Follow-up Observation Program website, which
is operated by the California Institute of Technology, under contract
with the National Aeronautics and Space Administration under the
Exoplanet Exploration Program. This paper includes data collected by
the \TESS\ mission, which are publicly available from the Multimission
Archive for Space Telescopes (MAST).  Some of the observations in this
paper made use of the High-Resolution Imaging instrument `Alopeke at
Gemini-North. `Alopeke was funded by the NASA Exoplanet Exploration
Program and built at the NASA Ames Research Center by Steve B. Howell,
Nic Scott, Elliott P. Horch, and Emmett Quigley.  Resources supporting
this work were provided by the NASA High-End Computing (HEC) Program
through the NASA Advanced Supercomputing (NAS) Division at Ames
Research Center for the production of the SPOC data products.
The authors wish to recognize and acknowledge the very significant
cultural role and reverence that the summit of Mauna Kea has always
had within the indigenous Hawaiian community. We are most fortunate to
have the opportunity to conduct observations from this mountain.

Data and materials
availability: All associated data products are available for download
from ExoFOP-TESS website,
\url{https://exofop.ipac.caltech.edu/tess/target.php?id=181804752}.

Facilities: \facility{TESS, Gaia, Gemini (`Alopeke), Keck I (HIRES),
  Keck II (NIRC2), Subaru (IRD), LCO }

\bibliographystyle{aasjournal}

\hspace{-1in}
\begin{deluxetable}{l l l }[bt]
\hspace{-1in}\tabletypesize{\scriptsize}
\tablecaption{  Stellar Parameters of \starname \label{tab:stellar}}
\tablewidth{0pt}
\tablehead{
\colhead{Parameter} & \colhead{Value} & \colhead{Source}
}
\startdata
\multicolumn{3}{l}{\hspace{1cm}\em Identifying information} \\
TIC ID & 181804752 & TIC v8 \citep{stassun:2018} \\
$\alpha$ R.A. (hh:mm:ss) & 11:02:45.96  &\\
$\delta$ Dec. (dd:mm:ss) & -16:24:22.29 & \\
$\mu_{\alpha}$ (mas~yr$^{-1}$) & -221.08 $\pm$ 0.22 & Gaia DR2 \\
$\mu_{\delta}$ (mas~yr$^{-1}$) & -59.00 $\pm$ 0.14 & Gaia DR2 \\
Distance (pc) & $26.493 \pm 0.064$ & Gaia DR2 \citep{bailerjones:2018}\\
\multicolumn{3}{l}{\hspace{1cm}\em Photometric Properties} \\
V (mag) .......... & 16.9 $\pm$ 0.2 & TIC v8  \\
G (mag) .......... & 15.0715 $\pm$ 0.0013 & Gaia DR2 \citep{gaia:2018}\\
G$_{BP}$ (mag) .......... & 17.23831 $\pm$ 0.0072 & Gaia DR2 \\
G$_{RP}$ (mag) .......... & 13.69512 $\pm$ 0.0029 & Gaia DR2 \\
u (mag)..........  &21.28  $\pm$ 0.14  & SDSS \citep{albareti:2017} \\
g (mag)..........  &17.8827  $\pm$ 0.0057  & SDSS \\
r (mag) .......... & 16.2672  $\pm$ 0.0039  & SDSS \\
i (mag)........... & 14.3142 $\pm$ 0.0035 & SDSS \\
z (mag)........... & 13.2565 $\pm$ 0.0035 & SDSS \\
J (mag)..........  & 11.559 $\pm$ 0.024 & 2MASS \citep{skrutskie:2006}\\
H (mag) .........  & 10.993 $\pm$ 0.022 & 2MASS\\
Ks (mag) ........  & 10.644 $\pm$ 0.023 & 2MASS\\
W1 (mag) ........  & 10.426 $\pm$ 0.023 & AllWISE \citep{cutri:2012}\\
W2 (mag) ........  & 10.233 $\pm$ 0.021 & AllWISE\\
W3 (mag) ........  & 10.024 $\pm$ 0.062 & AllWISE \\
\multicolumn{3}{l}{\hspace{1cm}\em Spectroscopic and Derived Properties} \\
Spectral Type & M(6.1$\pm$0.7)V & This work \\
Barycentric rv (km~s$^{-1}$) & +14.1 $\pm$ 0.3 & This work \\
Age (Gyr) & $>0.5$ & This work \\
$[$Fe/H$]$ & $-0.09 \pm 0.19$ & This work \\
$T_\mathrm{eff}$ (K) & 2960 $\pm$ 55 & This work \\
$\log_\mathrm{10} g$ (cgs) & 5.115 $\pm$ 0.094 & This work\\
$v \sin i$ (km~s$^{-1}$) & $<$2   &  This work\\
$M_*$ ($M_\odot$) &  0.139$\pm$0.005 & This work \\
$R_*$ ($R_\odot$) &  0.171$\pm$0.018 & This work \\
$L_*$ ($L_\odot$) &  0.00201$\pm$0.00045 & This work \\
\enddata
\end{deluxetable}


\begin{deluxetable*}{l l l l}[bt]
\tabletypesize{\scriptsize}
\tablecaption{  Planet Parameters \label{tab:planet}}
\tablewidth{0pt}
\tablehead{
\colhead{} & \colhead{} & \colhead{\starname b} & \colhead{\starname c} \\
\colhead{Parameter} & \colhead{Units} & \colhead{(TOI-736.02)} & \colhead{(TOI-736.01)}
}
\startdata
       $T_{0}$ & $BJD_\mathrm{TDB} - 2457000 $ & $1645.94405 \pm 0.00066$   & $1651.29807 \pm 0.00041$ \\
       $P$ &          d & $0.9480050 \pm 0.0000058$                           & $4.989963 \pm 0.000050$                     \\
       $i$ &        deg & $87.3^{+2.0}_{-4.9}$                                    & $89.55^{+0.32}_{-0.50}$                             \\
 $R_P/R_*$ &         -- & $0.0604\pm0.0028$                                   & $0.1238\pm 0.0022$                             \\
   $R_*/a$ &         -- & $0.090^{+0.058}_{-0.016}$                                & $0.0290^{+0.0035}_{-0.0016}$                        \\
\hline
  $T_{14}$ &         hr & $0.612^{+0.068}_{-0.079}$                                & $1.208^{+0.056}_{-0.046}$                           \\
  $T_{23}$ &         hr & $0.466^{+0.076}_{-0.259}$                                & $0.899^{+0.041}_{-0.048}$                           \\
       $b$ &         -- & $0.54^{+0.36}_{-0.37}$                                   & $0.28^{+0.24}_{-0.19}$                              \\
$\rho_{*,circ}$ & g~cm$^{-3}$ & $28\pm 22$                                     & $31.1^{+5.6}_{-9.1}$                          \\
       $a$ &         AU & $0.00969^{+0.00032}_{-0.00035}$                          & $0.029392^{+0.00098}_{-0.00105}$                     \\
     $R_P$ &      $R_E$ & $1.12\pm 0.13$                                & $2.31 \pm 0.25$                              \\
 $S_{inc}$ &      $S_E$ & $21.5^{+5.4}_{-4.6}$                                  & $2.35^{+0.59}_{-0.51}$                              \\
$T_\mathrm{eq}$\tablenotemark{a} & K & $650 \pm 120$ & $370 \pm 30$ \\
\enddata
\tablenotetext{a}{Assuming a uniform random distribution of Bond albedos (0--0.4) and heat redistribution factors (0.25--0.5).}
\end{deluxetable*}

\end{document}